\documentclass[prb,twocolumn,superscriptaddress,showpacs,floats,floatfix]{revtex4}
\usepackage{graphicx}
\usepackage{dcolumn}
\usepackage{bm}

\begin{document}
\title{Origin of the quasi-universality of the graphene minimal conductivity}

\author{J. J. Palacios}
\affiliation{Departamento de F\'{\i}sica de la Materia Condensada, Universidad Aut\'onoma de Madrid, 
Cantoblanco, Madrid 28049, Spain}

\begin{abstract}

It is a fact that the minimal conductivity $\sigma_0$ of most graphene samples is larger than the 
well-established universal value for ideal graphene $4e^2/\pi h$; in particular, 
larger by a factor $\gtrsim\pi$.  Despite intense theoretical activity, this fundamental issue has eluded an
explanation so far.  Here we present fully  atomistic quantum mechanical estimates of the graphene minimal 
conductivity where electron-electron interactions are 
considered in the framework of density functional theory. We show the first conclusive evidence of the dominant role
on the minimal conductivity of charged impurities over ripples, which have no visible effect. Furthermore, 
in combination with the logarithmic scaling law for diffusive metallic graphene, we ellucidate 
the origin of the ubiquitously observed minimal conductivity in the range $8e^2/h > \sigma_0 \gtrsim 4e^2/h$.
\end{abstract}


\maketitle

\section{Introduction}
The ensuing years after the first reported transport measurements of isolated graphene flakes have witnessed an intense
debate on the origin of the {\em quasi}-universal value of the minimal conductivity $\sigma_0$. 
The fact that experimental confirmations of the celebrated universal minimal conductivity 
$4e^2/\pi h$, expected for ideal  graphene, are the 
exception\cite{Miao07,PhysRevLett.100.196802} rather than the rule 
and that most graphene samples present larger values, typically
$\gtrsim 4e^2/h$, regardless of varying experimental 
conditions\cite{Novoselov05,Geim.Nat.Mat.07,PhysRevLett.99.246803,PhysRevLett.102.206603,Jang08-1,Chen08},
remains one the major fundamental unresolved questions in graphene physics. Although this issue
has spurred a vast amount of theoretical work\cite{bardarson:106801,PhysRevB.76.195445,PhysRevB.79.075405,lewenkopf:081410,PhysRevLett.102.106401,Nomura07,Adam07,PhysRevB.79.245423,PhysRevB.78.115426,PhysRevB.79.201404,PhysRevB.81.121408,PhysRevB.79.184205,PhysRevB.76.205423,fogler:236801}, 
to date, the answer remains elusive and no theory has been able to render a full picture of this jigsaw puzzle.

Numerical\cite{bardarson:106801,PhysRevB.76.195445} as well as analytical work\cite{PhysRevB.79.075405} 
for simple models of disorder and non-interacting electrons indicate that, as long as intervalley scattering is avoided, 
single-parameter scaling applies. This means that a beta function 
$\beta(\sigma_0)=d{\rm Ln}(\sigma_0)/d{\rm Ln}(L)$ exists, 
where $L$ is the size of the sample. In particular, $\beta(\sigma_0)>0$,
behaving as $1/(\pi\sigma_0)$ for $\sigma_0 \rightarrow \infty$.
In other words, neutral graphene would not present a metal-insulator transition, 
being its conductivity  bounded from below by the universal value 
$\sigma_0=4e^2/\pi h $ for ideal graphene
and unbounded from above ($\sigma_0 \rightarrow \infty$) as the disorder is increased.
This prediction ultimately relies on the expectation that intervalley scattering is not 
activated for most common types of (long-ranged) disorder and relevant length scales, 
a fact confirmed by some atomistic simulations\cite{lewenkopf:081410}, 
but questioned by others\cite{PhysRevLett.102.106401}.  

Alternatively, there have been efforts to consider likely sources of disorder such as nearby charged 
impurities\cite{Nomura07,Adam07,PhysRevB.79.245423,PhysRevB.78.115426}, ripples\cite{PhysRevB.79.184205}, and
resonant scatterers\cite{PhysRevB.76.205423} in the most possible realistic manner. 
This usually comes at the expense of a rigourous quantum mechanical treatment of transport which renders
$\sigma_0$ size independent\cite{Adam07,PhysRevB.79.245423,PhysRevB.76.205423,fogler:236801,PhysRevB.79.184205}. 
A number of predictions can be made in these semiclassical aproaches, having in common, e.g., the fact that
$\sigma_0$ always decreases with increasing impurity concentration 
$n_{\rm imp}$ in addition to not being bounded from above 
as $n_{\rm imp} \rightarrow 0$. This is at odds with the theoretical predictions mentioned in the previous paragraph
and hardly agrees with available experimental 
evidence\cite{Novoselov05,Geim.Nat.Mat.07,PhysRevLett.99.246803,PhysRevLett.102.206603,Jang08-1,Chen08}.
The inherent limitations of semiclassical approaches have been recently
appreciated\cite{PhysRevB.79.201404,PhysRevB.81.121408} and are at the heart of the discrepancy.  
One cannot deny, however, the insight gained on the actual microscopic origin of the minimal conductivity from
a realistic treatment of disorder. 
Combining a full quantum mechanical approach to transport with
a realistic description of disorder and screening seems
to be the only way to resolve the controversy and this is our major contribution in this work.  

Our findings can be summarized as follows. (i) First,
our calculations corroborate previous theoretical 
work and agree with experimental findings such as the observed linear behavior of the conductivity with 
electron density as well as the observed different mobilities for electrons and holes.
(ii) Second, charged impurities prevail over ripples in the experimentally relevant
range of impurity concentrations and distances to the graphene flake, increasing the clean-limit conductivity.
(v) Finally, this increase is limited to a small percentage of the clean-limit value for small systems, 
which, in combination with
the logarithmic scaling law predicted in the absence of intervalley scattering, allows us to explain 
the narrow range of values $8e^2/h > \sigma_0 \gtrsim 4e^2/h$ of the observed graphene minimal conductivity.

\begin{figure}
\includegraphics[width=1.0\linewidth]{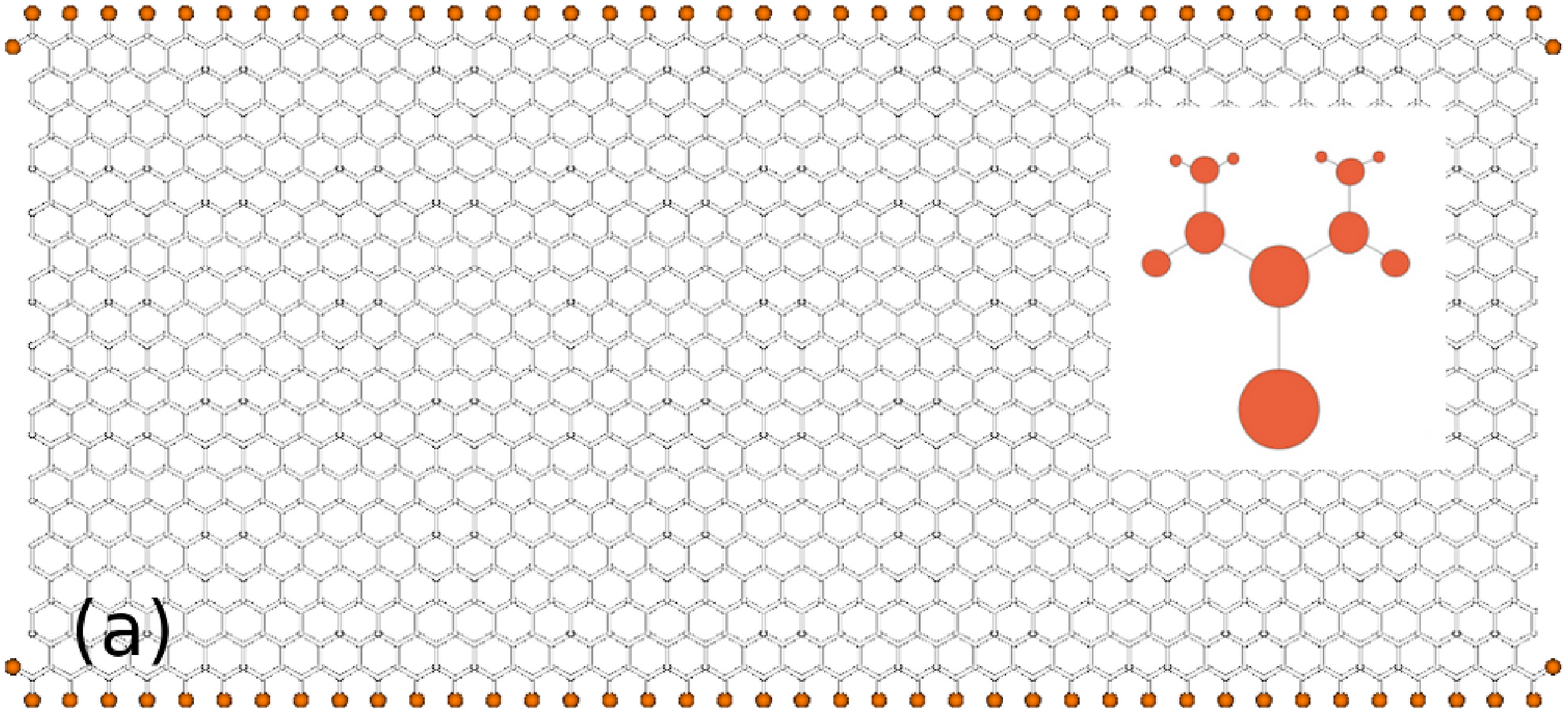}
\begin{tabular}{cc}
\includegraphics[width=0.4\linewidth]{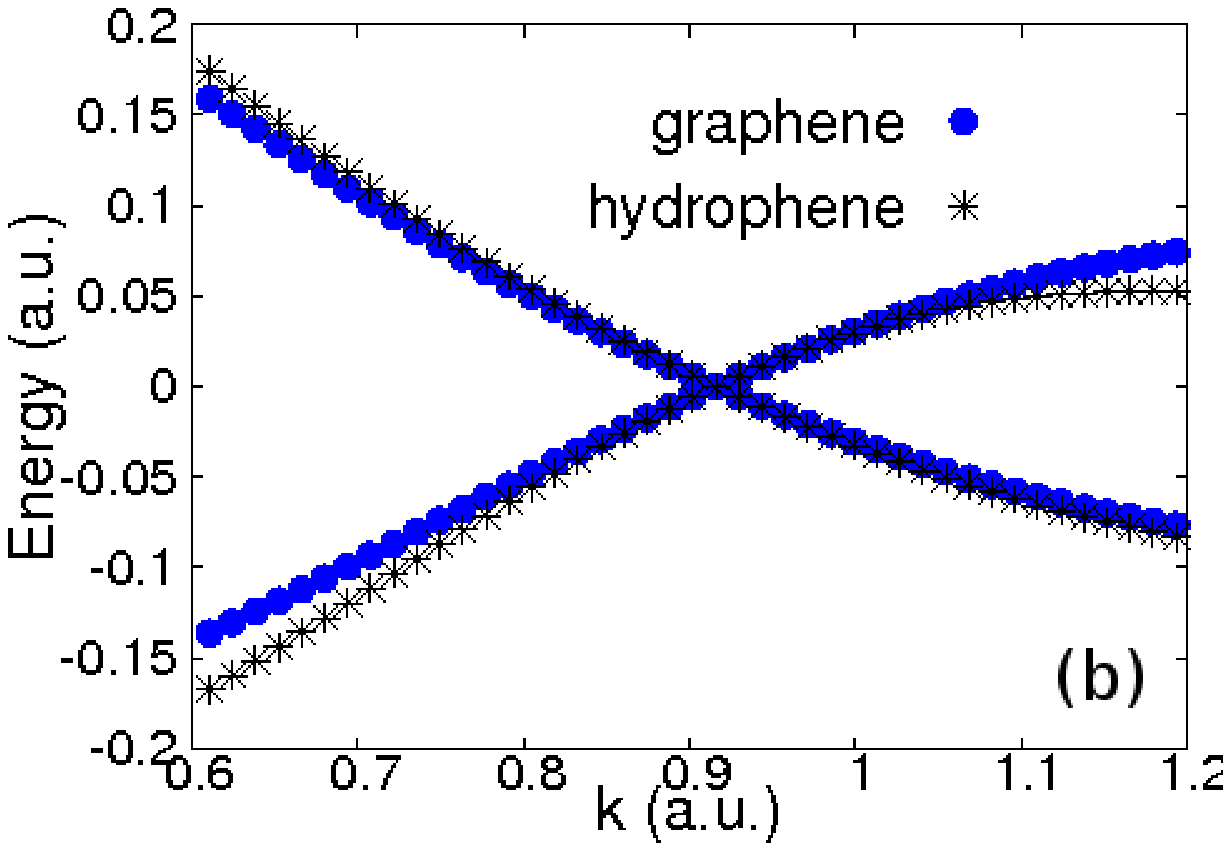}&
\includegraphics[width=0.4\linewidth]{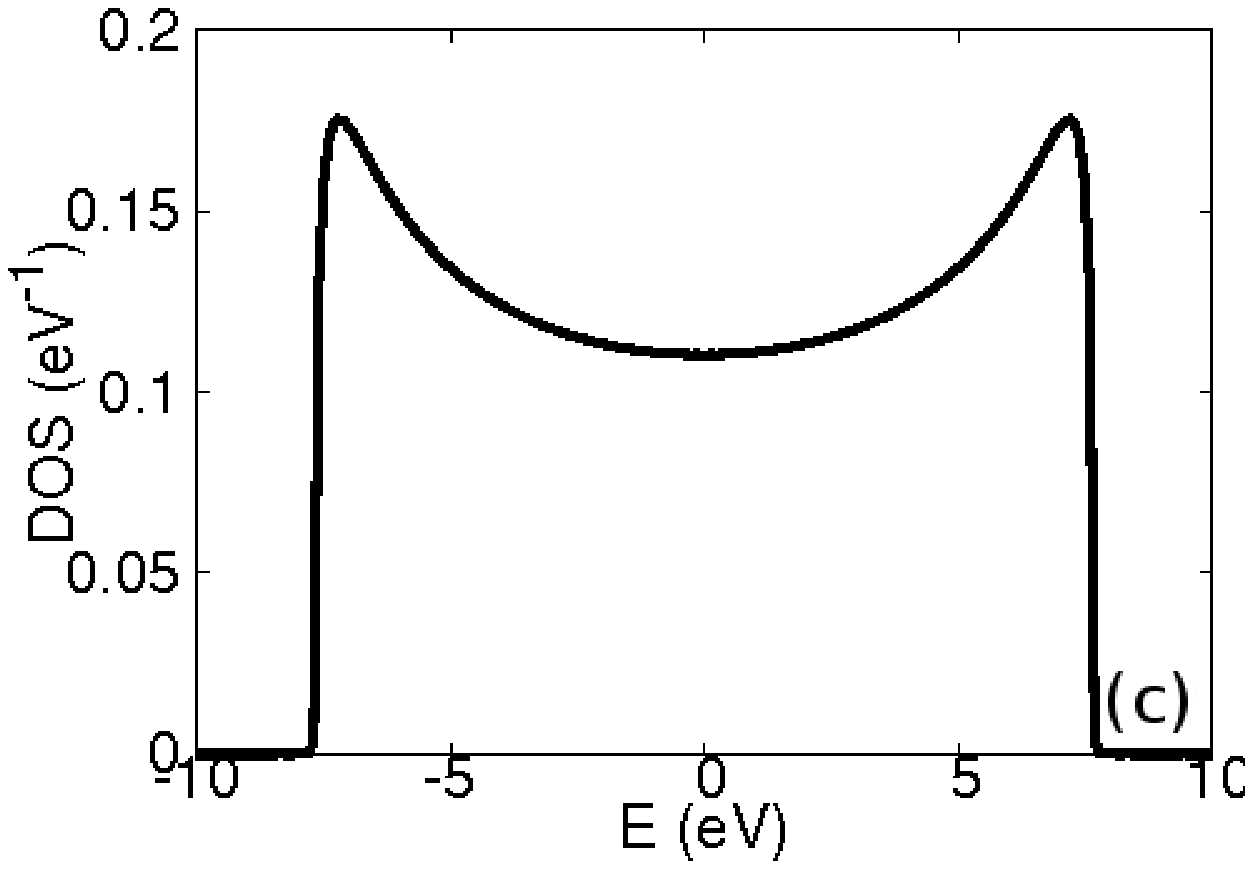}
\end{tabular}
  \caption{Color online. (a)
A hydrophene ribbon contacted to metallic electrodes as to drive the current in the armchair direction
(only the first layer of the electrodes is shown). The
electrodes are modeled here by a bidimensional three-fold coordinated tight-binding Bethe lattice, schematically
shown in the inset.  (b) Band structure (in atomic units) in the 
local density approximation near one of the  Dirac points for graphene and hydrophene where  
a minimal basis set has been used for graphene.  (c)
Electrodes density of states per atom for a nearest-neighbors hopping amplitude $t= 2.7$ eV.}
\label{preliminar}
\end{figure}

\section{Hydrophene: A minimal model}
We consider a model for graphene where the C atoms are replaced by hydrogenic atoms. 
The $\pi$ orbitals of the C atoms, responsible for the low
energy physics,  are represented by the $s$ orbitals of the hydrogenic atoms, which, in turn,
are approximated by a gaussian function.
The gaussian exponent is optimized to reproduce the low-energy band structure of real graphene when both are computed in 
the local density approximation (LDA) as implemented in GAUSSIAN03\cite{GAUSSIAN:03}
[see Fig. \ref{preliminar}(b)].  We will refer to this model as {\em hydrophene}. We are interested in the conductivity,
defined through $\sigma=G\frac{L}{W}$, where $G$ is the conductance
of finite hydrophene ribbons of width $W$ along the zigzag direction ($x$) and length $L$ along the armchair direction
($y$). The latter coincides here with the direction of the current  which is driven by metallic
electrodes contacted along the width of the ribbons as shown in Fig. \ref{preliminar}(a).
The metallic electrodes are modeled by a bidimensional tight-binding Bethe lattice
of coordination three and only nearest-neighbors hopping
where the six crystallographic directions coincide with those of the ribbon (see inset). This electrode model relates
closely to graphene, but presents a finite density of states at the neutrality point $E=0$
[see Fig. \ref{preliminar}(c)]. As shown in
Fig. \ref{preliminar}(a), a branch of the Bethe lattice is connected to each undercoordinated atom on the zigzag edge.
 The minimal conductivity (defined in the limit 
$W/L \rightarrow \infty$) depends on the particular choice for the electrode model, but never exceeds the universal value 
$\sigma_0=\frac{4e^2}{\pi h}$\cite{Tworzydlo06,brey:116802} for ideal graphene.
We have chosen here metallic ribbons, i.e., those that, when their electronic structure is computed
at the simplest nearest-neighbors tight-binding model, this does not present a gap for $L\rightarrow \infty$.
The results do not depend either on this choice or the direction of the injected current.   

\section{possible sources of disorder}
We consider here two types of disorder currently accepted to possibly influence
the mobility and the minimal conductivity of graphene: Charged impurities and ripples. 

\begin{figure}
\includegraphics[width=1.0\linewidth]{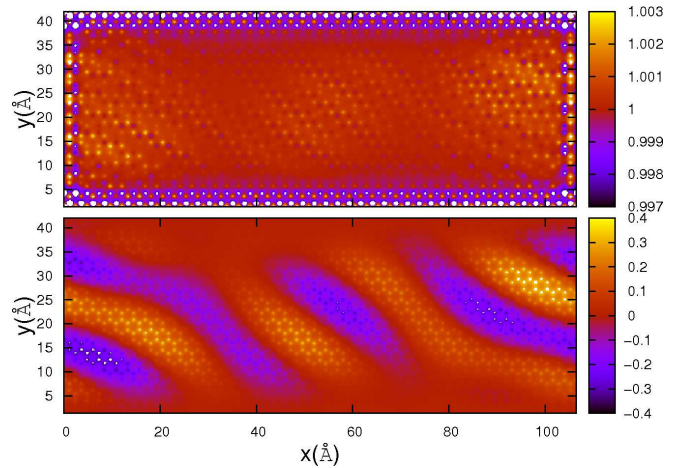}
  \caption{Color online. Upper panel shows an example of the induced charge density due to the 
ripples shown in the lower panel.
The scale in the upper panel refers to charge per atom which has been smoothed over a fine grid. The scale in the lower
panel is given in \AA.}
  \label{ripples}
\end{figure}

The model for the ripples consists of a randomly generated position-dependent height function given by the equation
\begin{equation}
h(x,y)=\sum_{\vec{k}}^{N_k} \frac{A\pi}{N_k|\vec{k}|} \sin{(k_xx+k_yy)},
\end{equation}
where $N_k$ is the number of Fourier components ($\approx 5$) and $A$ is 
the parameter that accounts for the overall deviation from planarity which 
is chosen as to reproduce typical ripple height-to-size ratios in the order of $\approx 0.1$.
$2\pi/W < k_x\ll 2\pi/a$ and $2\pi/L < k_y\ll 2\pi/a$ are random reciprocal wave vectors where $a$ is the graphene
lattice constant. Since in the hydrophene model there is only one
spherical orbital per site, changes in height only affect the hopping between atoms (and the overlap), creating
an accompanying pseudo-magnetic field landscape\cite{PhysRevB.81.035408} which does not affect the charge
density distribution. To account for changes in the on-site energies due to the $sp^3$ rehybridization 
in real graphene, we have considered a scalar potential 
of the type $\phi(x,y)=-B\left(\nabla^2 h(x,y)\right)^2$\cite{Kim08}, where $B>0$ is chosen as to reproduce potential
fluctuations in the order of a few tens of meV\cite{Kim08}. An example of the corrugation and LDA induced charge is shown 
in Fig. \ref{ripples}.

\begin{figure}
\includegraphics[width=1.0\linewidth]{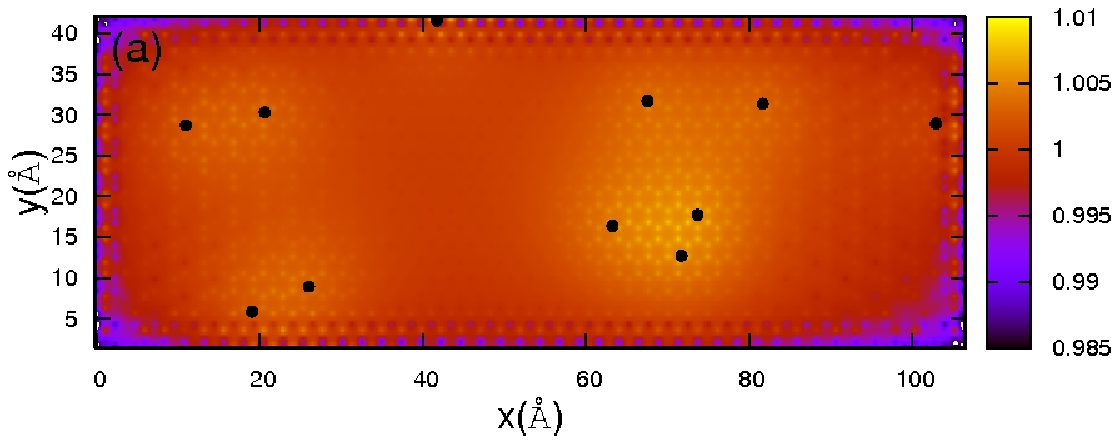}
\includegraphics[width=1.0\linewidth]{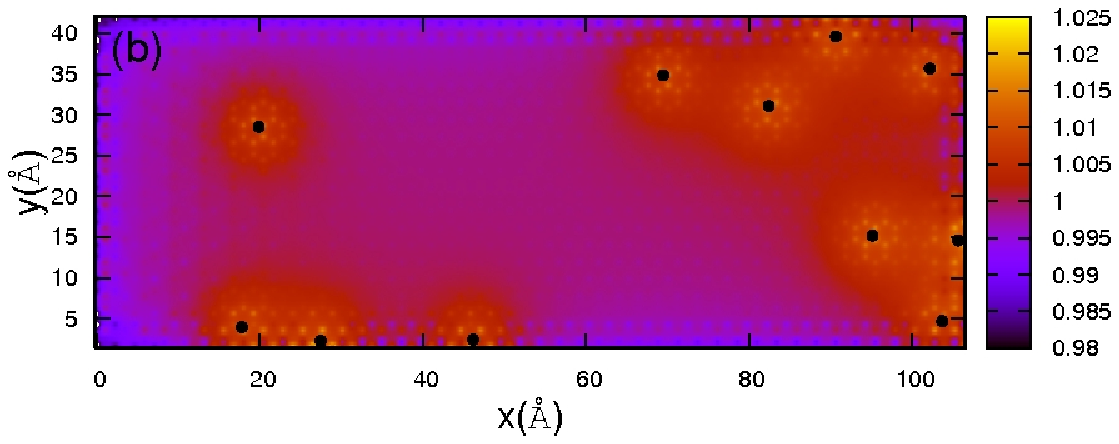}
\includegraphics[width=1.0\linewidth]{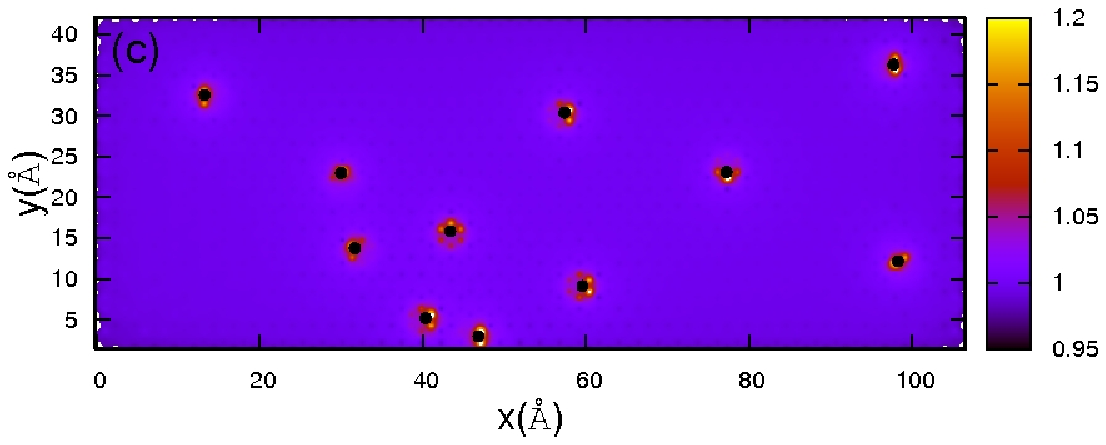}
  \caption{Color online. Screening electron density induced by charged impurities 
(black dots) placed at a distance $D=10$ (a), $D=5$ (b), and
$D=1$ (c) \AA.  Overall charge neutrality within the graphene flake has been imposed in all cases.
The scale refers to charge per atom which has been smoothed over a fine grid.}
  \label{density}
\end{figure}

\begin{figure}
\includegraphics[width=1.0\linewidth]{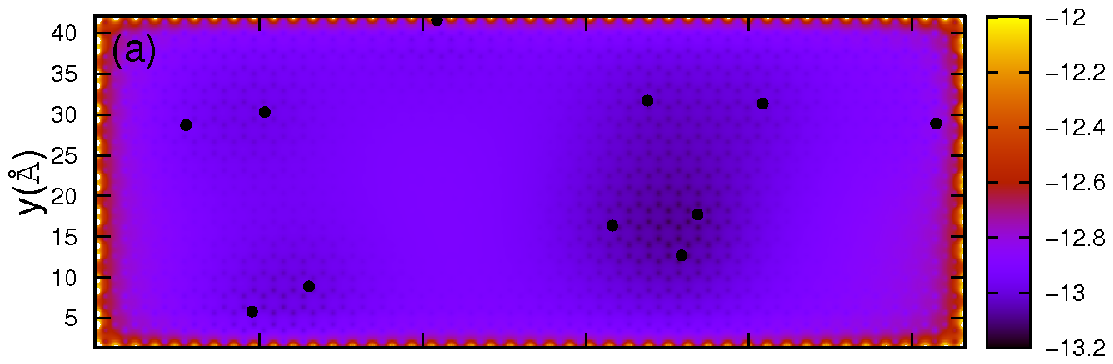}
\includegraphics[width=1.0\linewidth]{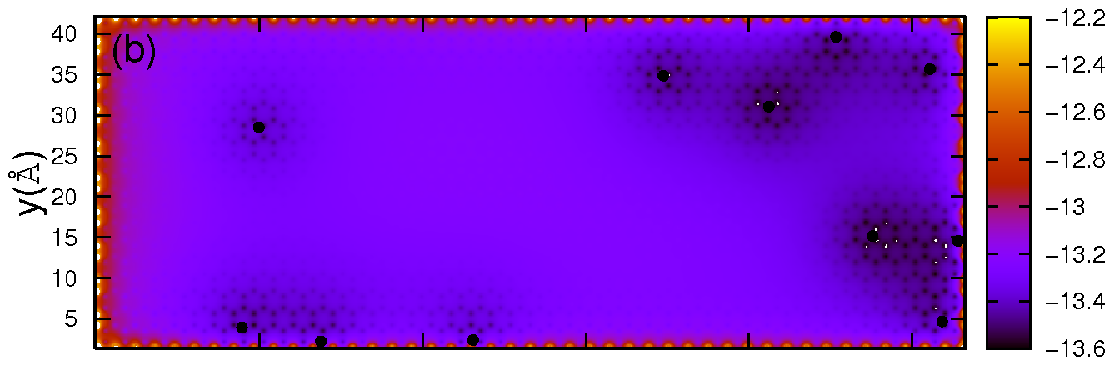}
\includegraphics[width=1.0\linewidth]{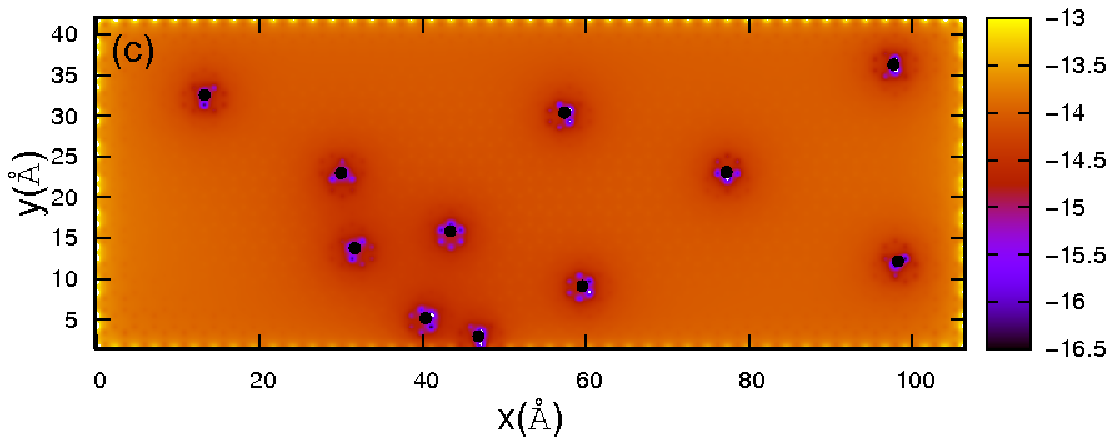}
  \caption{Color online. Kohn-Sham potential in the graphene plane as induced by charged impurities (black dots) placed 
at a distance $D=10$ (a), $D=5$ (b), and $D=1$ (c) \AA $\:$ off the graphene plane. The scale is in eV.
Overall charge neutrality has been imposed in all cases within the graphene flake.}
  \label{potential}
\end{figure}

For simplicity in the analysis of the results, we consider $Z=1$ impurities randomly distributed
in the same plane at a distance $D$ off the hydrophene ribbon. This is probably a good assumption since charged
impurities are expected to be located near the surface of the SiO$_2$ substrate or in between the surface and the
graphene flake or even adsorbed on the graphene flake, depending on their specific origin. 
Figure \ref{density} shows the screening charge density for various representative 
disorder realizations of charged impurities at different distances off the graphene plane. 
Starting at $D=10$ \AA, a landscape of electron-hole puddles gives way to strongly localized screening charge  around
each impurity as $D\rightarrow 0$. 
This screening charge approaches the $\delta-$function predicted by effective models\cite{Shytov07},
although regularized by the finite value of $D$ and by the lattice constant of our atomistic model.
Figure \ref{potential} shows the accompanying Kohn-Sham potential.
The range of the potential induced by a single
impurity can be as small as $\xi \approx 0.5$ nm and the depth as high as 2 eV for $D=1$ \AA.
which, in principle, can induce strong intervalley scattering\cite{PhysRevLett.102.106401}.
Electron-hole puddles are also induced by the ripples due to changes in second-neighbor hopping, 
rehybridization, and local changes in exchange interactions\cite{Brey09}, but the average induced electronic density is
still typically one order of magnitude smaller than the one induced by impurities at $D=10$ \AA. 


\section{conductivity results}
For the conductivity calculations we have employed the code ANT.G03, which is part of
the quantum transport computational toolbox ALACANT\cite{palacios:prb:01,palacios:prb:02,ALACANT}.  
The basics of the calculation are as follows:
i) The Bethe lattice is incorporated into the partially pre-computed Green's function 
of the isolated hydrophene ribbon, $G_H$, in the form of a self-energy $\Sigma$, ii) 
a density matrix is obtained from the new Green's function imposing overall charge neutrality in the ribbon, 
iii) a new Green's function is evaluated from the previously obtained density matrix, 
and iv) the procedure is repeated until a self-consistent solution is reached.
For the LDA exchange correlation functional we have employed the standard approximation as implemented in 
GAUSSIAN03\cite{GAUSSIAN:03}.
With the self-consistent Green's function and the self-energies of the left($\Sigma_L$) and right($\Sigma_R$) 
electrodes, the conductance can be calculated from the Landauer formalism:
\begin{equation}
G(E)=\frac{2e^2}{h}{\rm Tr}\left[{G}_H^{\dagger}(E){\Gamma}_R(E){G}_H(E){\Gamma}_L(E)\right],
\label{conductance}
\end{equation}
where ${\Gamma}_{R(L)}=i\left({\Sigma}_{R(L)}-{\Sigma}_{R(L)}^{\dagger}\right)$. 

\begin{figure}
\includegraphics[width=1.0\linewidth]{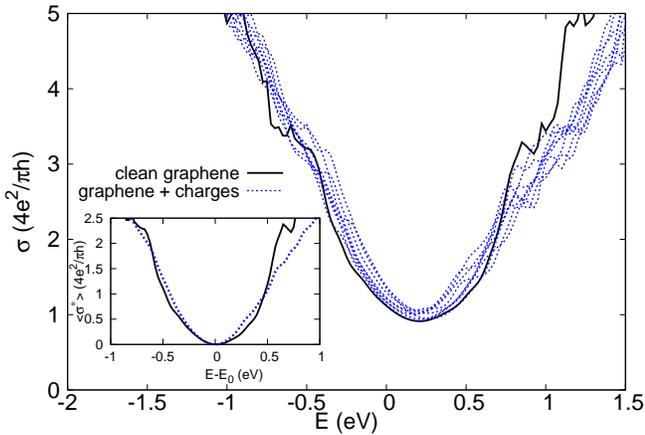} 
\caption{Color online. Conductivity as a function of energy.
The result for a clean ribbon with aspect ratio $W/L\approx 3$ is shown  by the black solid line. The results for various
realizations of randomly distributed impurities at a distance $D=1$ \AA $\:$
corresponding to a density of $n_{\rm imp}\approx 2.0 \; 10^{13}$  cm$^{-2}$ are shown by blue
dashed lines. Inset: Same as in main plot, but after
averaging over disorder realizations. The minima have been offset to the origin here.} 
\label{sigma}
\end{figure}

\begin{figure}
\includegraphics[width=1.0\linewidth]{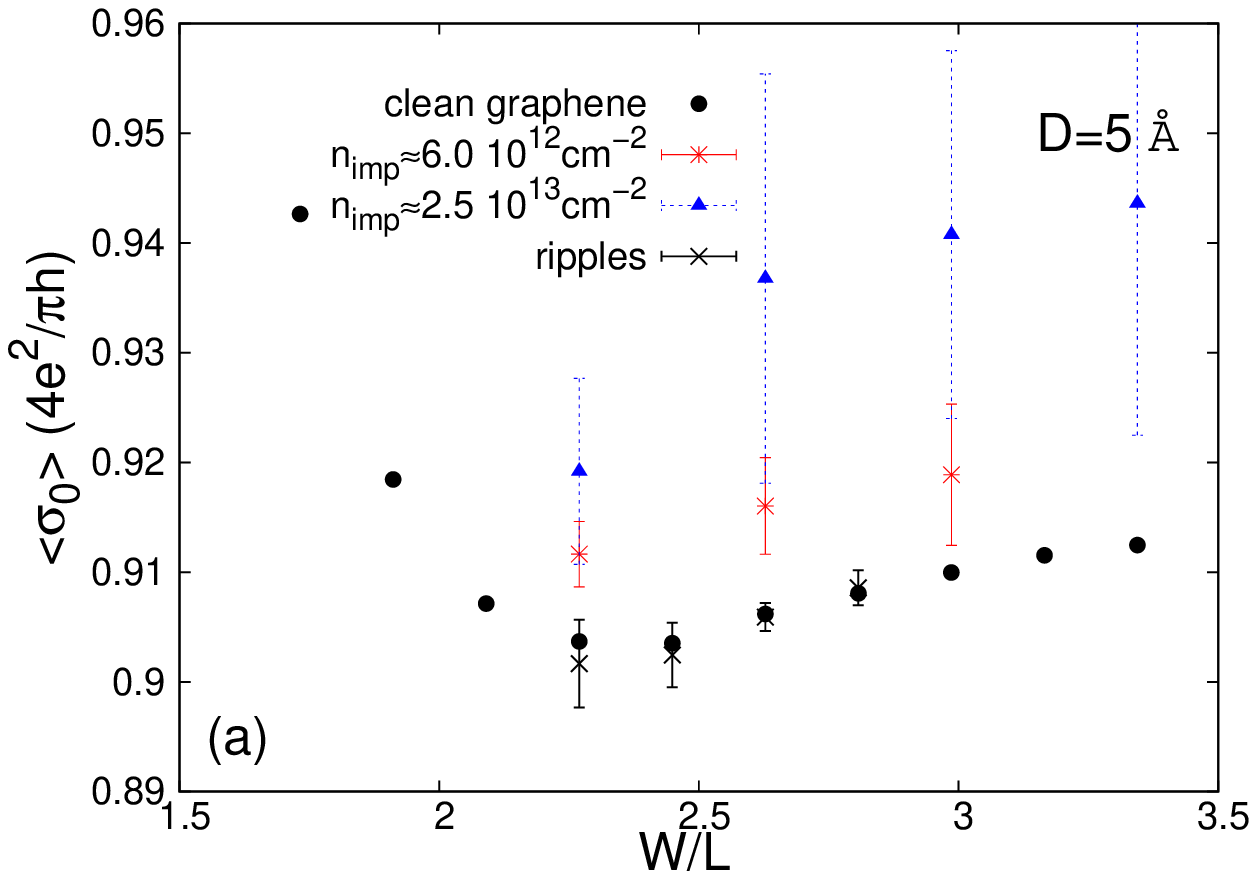} 
\includegraphics[width=1.0\linewidth]{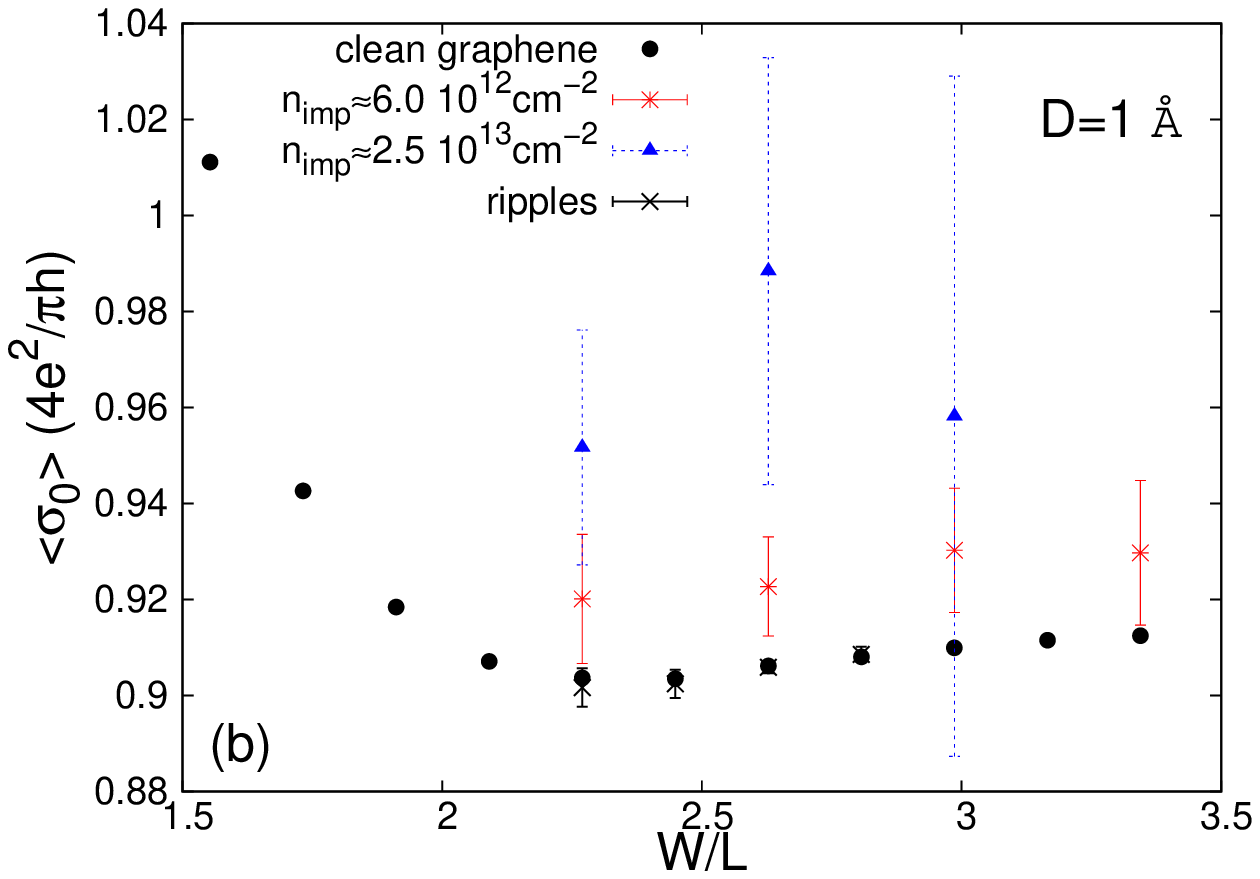}
\caption{Color online. Conductivity as a function of the aspect ratio. (a) The 
results for flat and clean graphene are represented
by black dots and the ones including only ripples by crosses. 
The results for charged impurities placed at a distance $D=5$ \AA $\:$ 
off the graphene plane are shown by triangles and stars for two different concentrations. (b) Same as in (a) but
for a distance $D=1$ \AA. }
  \label{sigmamean}
 \end{figure}

The LDA conductivity $\sigma$ as a function of energy 
is presented in Fig. \ref{sigma} for a ribbon with an aspect ratio $W/L \approx 3$.
The length here is $L\approx 4$ nm and will remain the same in all the calculations. 
The solid line corresponds to the ideal or ``clean'' case with neither ripples nor impurities.
The Dirac point $E_0 $ appears shifted to positive energies (the Fermi energy has been set to zero)
due to the influence of the metallic electrodes, despite the fact that overall charge neutrality is imposed on the ribbon. 
We have checked, by shifting the Fermi energy upwards to the Dirac point $E_0$ on electron doping, that
the overall conductivity curve is not appreciably affected with respect to the undoped case. From now on we will
take $\sigma_0=\sigma(E_0)$ (for $E_0$ typically $>0$) as the minimal conductivity.
This is now shown as a function of $W/L$
(black dots) in Fig. \ref{sigmamean}.  The fact that $\sigma_0$ scales with $W/L$ to a 
value slightly smaller than $4e^2/\pi h$ for $W/L \rightarrow \infty$\cite{Tworzydlo06}
can, in principle, be attributed to the chosen electrode model. 
One could possibly improve this result, i.e., increase the conductivity closer to the
universal value $4e^2/\pi h$, by tuning the tight-binding parameters of the Bethe lattice or by using
a different model for the electrodes\cite{Tworzydlo06,brey:116802}. 
Whether or not the LDA minimal conductivity of clean graphene
can reach the universal value $4e^2/\pi h$ is, anyhow, not essential in the ensuing discussion.
 
We now turn our discussion to the effect of charged impurities and ripples on the conductivity.
Figure \ref{sigma} shows $\sigma(E)$  for various realizations of randomly distributed
impurities at fixed $n_{\rm imp}$ and $D$ (dashed lines).  Defining $\sigma^*(E)=\sigma(E+E_0)-\sigma_0$ 
and averaging over impurity realizations,
it becomes apparent (see inset) 
that $\langle\sigma^*\rangle\propto E^\alpha$ with $\alpha\approx 1$ (more visible for electrons)
in contrast to the clean limit case where $\alpha > 1$. This result, in addition to our numerical evidence that
the chemical potential depends linearly on the density  for high values of $n_{\rm imp}$ (see also
Ref. \onlinecite{PhysRevLett.101.166803}),
provides further confirmation that $\sigma(n) \propto n$, as experimentally observed and previously explained
in the Boltzmann transport approximation\cite{Nomura07,Adam07,PhysRevB.79.245423}.
It also becomes apparent in the inset that the mobility ($\mu=\sigma/ne$) of electrons decreases as 
compared to that of holes\cite{novikov:102102}. 
All these results are nicely compatible with previous works and give us confidence on
the validity of our LDA results for hydrophene.

\begin{figure}
\includegraphics[width=1.0\linewidth]{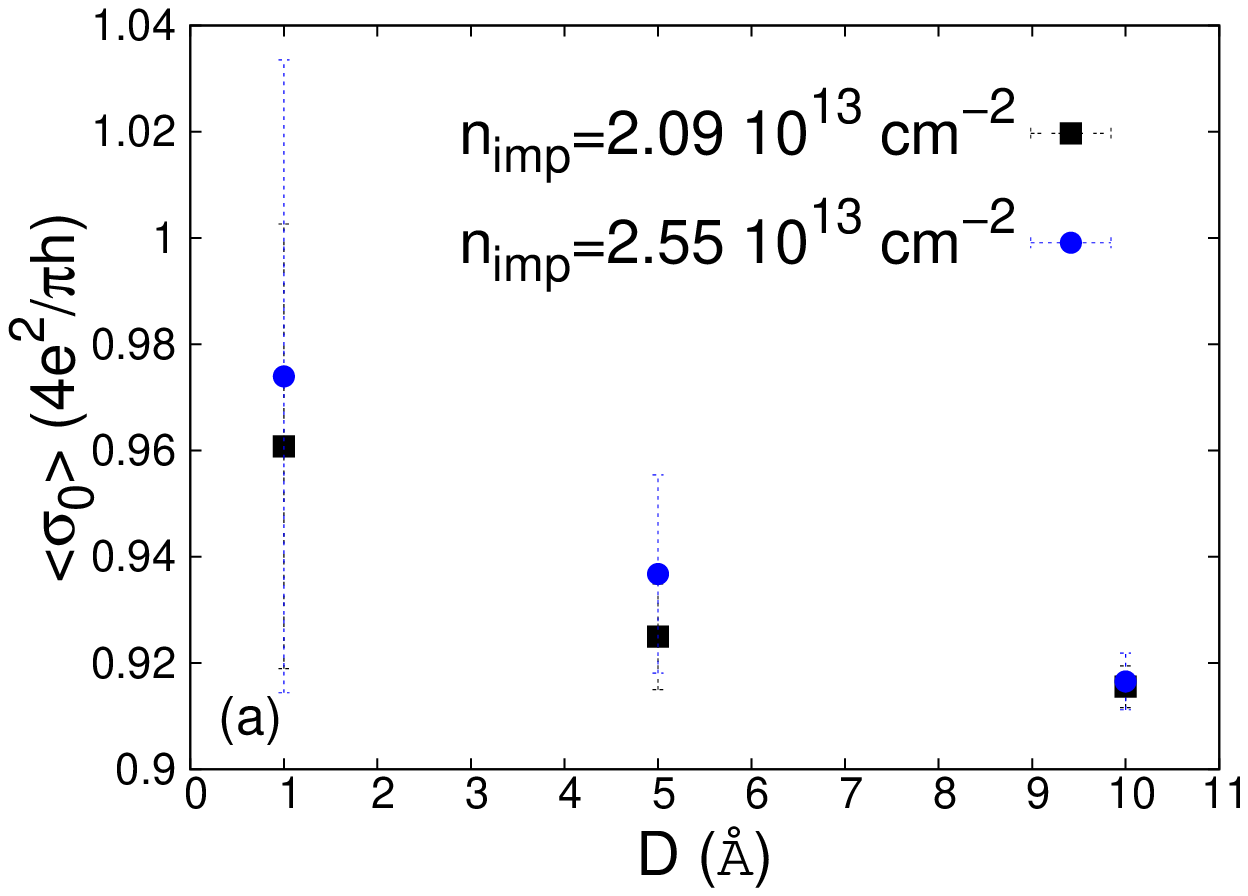} 
\includegraphics[width=1.0\linewidth]{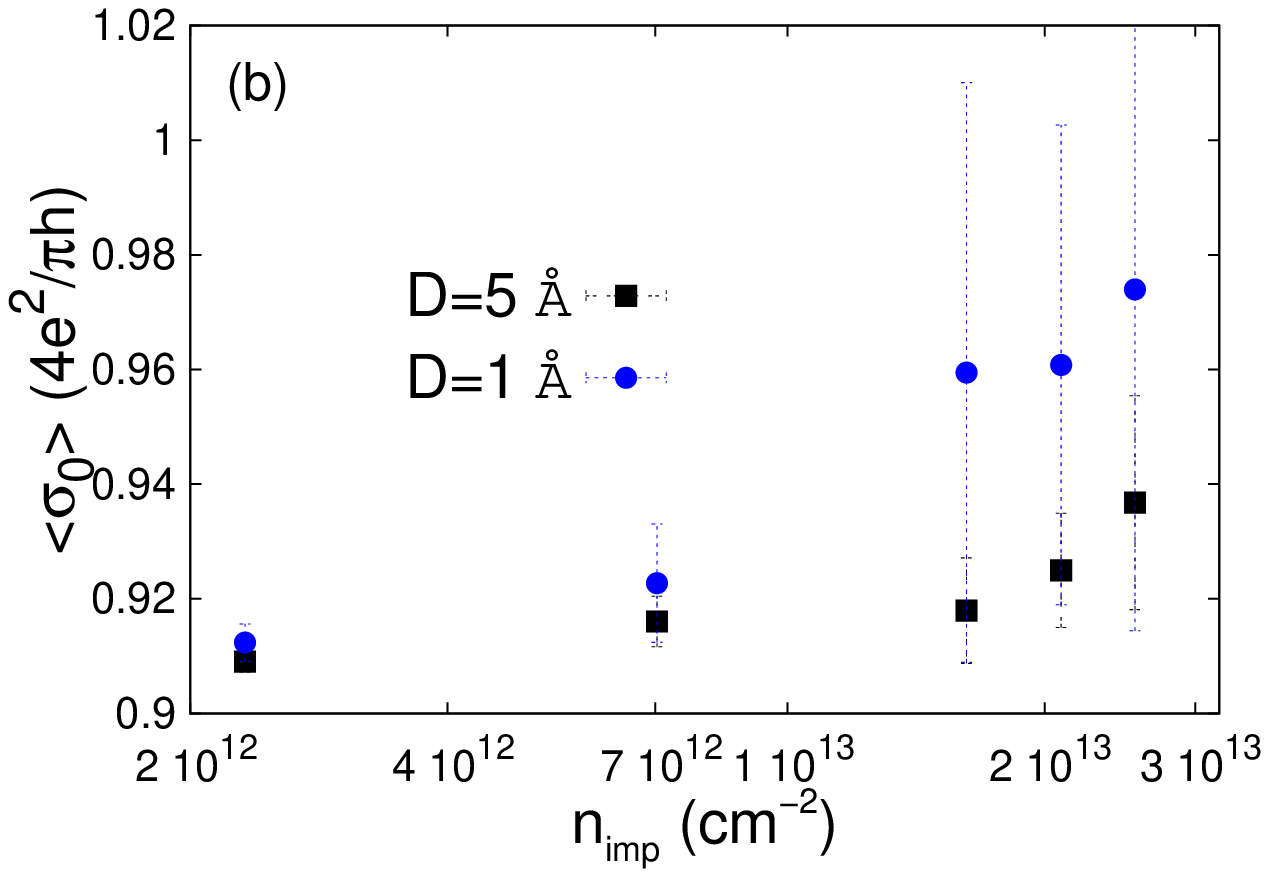}
  \caption{Color online. Dependence of the minimal conductivity on disorder for
a representative ribbon with an aspect ratio of $W/L\approx 2.6$. (a) Results
for two different concentrations of charged impurities as a function of their distance to the graphene plane. (b) 
Results for two different distances as a function of the concentration of impurities.  }
  \label{moresigma}
\end{figure}

On top of the clean graphene minimal conductivity,
Fig. \ref{sigmamean} also shows $\langle\sigma_0\rangle$ for a large set of impurity
and corrugated graphene realizations for two values of $D$ and two values of $n_{\rm imp}$. 
Each point has been obtained after averaging over 15-20 realizations. 
The results clearly indicate that,  for the chosen range of $D$ and $n_{\rm imp}$,
the latter typically one order of magnitude larger than the value estimated 
for exfoliated graphene deposited on SiO$_2$ ($\approx 10^{11}-10^{12} $ cm$^2$), 
charged impurities {\em increase} the minimal conductivity. 
Ripples, if anything, add some dispersion to the clean-limit values 
but their influence is negligible compared to that of charges, at least in the range of parameters shown. 
Importantly, the unavoidable and large statistical uncertainty
does not withstand the fact that $\langle \sigma_0 \rangle$ scales to a constant value with $W/L$, 
allowing us to define a minimal conductivity for a given length $L$. 
A downward deviation is, nevertheless, apparent at large $W/L$ for strong disorder ($D=1$ \AA)
which signals the activation of localization.  For completeness,
we also present a systematic study of $\langle\sigma_0\rangle$ for  $W/L\approx 2.6$.
$\langle\sigma_0\rangle$ increases as $D$ decreases  
down to the smallest meaningful value, $D=1$ \AA $\:$ [see Fig. \ref{moresigma}(a)],
and increases with increasing $n_{\rm imp}$ up to the highest value considered
$n_{\rm imp} \approx 2.5 \: 10^{13}$ cm$^{-2}$ [see Fig. \ref{moresigma}(b)].

A scaling analysis with $L$ is beyond present numerical capabilities for self-consistent calculations. Nevertheless, 
an increasing conductivity with increasing disorder strength is compatible with the 
scaling law for diffusive metallic graphene in the absence of 
intervalley scattering\cite{bardarson:106801,PhysRevB.79.075405,PhysRevB.78.235438}:
\begin{equation}
\sigma_0/(4e^2/h)= \frac{1}{\pi}{\rm Ln}(L/l_{\rm intra}),  
\label{scaling}
\end{equation}
where $l_{\rm intra}$ can be interpreted as 
the disorder-dependent intra-valley mean free path which, in principle, can be estimated from our numerics. 
 Notice that regardless of whether or not the weak disorder cases (large $D$ and/or small $n_{\rm imp}$)
lie outside the realm of the diffusive regime for the system sizes considered, 
Eq. \ref{scaling} is certainly valid for the strong disorder case close to the critical value of 
 $W/L$ beyond which inter-valley scattering decreases the conductivity
[see Fig. \ref{sigmamean}(b)]. The key observation is now that $\sigma_0$ is only increased by as much as 
$\approx 10$\% with respect to the clean limit value
(see Figs. \ref{sigmamean} and \ref{moresigma}) before localization sets in. This sets a minimum effective 
value for $l_{\rm intra}$ of $\approx 1.5$nm with a very weak dependence on disorder over approximately a decade of 
(smaller) impurity concentrations and (larger) distances of the impurities to the graphene flake\cite{footnote}.
This minimum length can be used now to estimate the {\em maximum} possible value of the minimal conductivity
for the experimentally largest system sizes, $\sigma_0^{\rm max}(L\approx1\mu{\rm m})\approx 8e^2/h$. (For larger samples
$L$ may always be effectively limited by the phase coherence length $l_\phi \lesssim 1\mu$m\cite{Ki08-1}.)
In summary, {\em the weak logarithmic $L$-dependence of $\sigma_0$ in Eq. \ref{scaling} and the effective minimum
intra-valley mean free path in the order of $\approx$ 1.5 nm 
combine to approximately cancel the factor $\pi$ for relevant length scales ($0.1 \mu$m$\: < L,l_\phi <  1 \mu$m)
and explain why most graphene samples exhibit minimal conductivity values in the range $8e^2/h > \sigma_0 \gtrsim 4e^2/h$.} 

A few final remarks are in order. 
(i) For strong disorder or very large samples with unintentional disorder, i.e., 
for $l_{\rm inter}\ll L$, where $l_{\rm inter}$ is the inter-valley scattering length,
graphene behaves as an insulator.  The former condition may be achieved by intentional doping\cite{Chen08}.
The latter, however, may be prevented by the temperature-dependent $l_\phi$, which effectively
limits $L$ to $\lesssim 1 \mu$m\cite{Ki08-1}. (ii)
When $l_{intra} > L$, for instance, for very weak disorder or for very short samples with unintentional disorder, 
the scaling law in Eq. \ref{scaling} for diffusive systems does no longer apply 
and  $\sigma_0$ approaches the universal value $4e^2/\pi h$ as, e.g.,
reported in Refs. \onlinecite{Miao07,PhysRevLett.100.196802}.
(iii) Screening from the substrate or the occupied bands (not included in the model) 
can only increase $l_{\rm intra}$, strengthening  the quasi-universal character of
$\sigma_0$. (iv) Finally, when zero-energy states (or resonant states) appear due to the presence
of covalently-bonded adsorbates or vacancies, a number of predictions that range from the expectation of an
insulating behaviour\cite{Robinson08} at relevant scales to a finite minimal
conductivity\cite{PhysRevB.76.205423} have been put forward. These, certainly, might play a role
in the mobility\cite{Ni.1003.0202v2} and minimal conductivity of graphene, but this study is out the scope of
this work.

In summary, we have evaluated the conductivity of graphene in the presence of charge impurities and ripples including
the screening in the local density approximation. Impurities are solely 
responsible for the increase of the minimal conductivity with respect to the clean-limit universal value. We have
quantified this increase and estimated that the minimal conductivity normally lies in the range
$8e^2/h > \sigma_0 \gtrsim 4e^2/h$ as observed in experiments.

\begin{acknowledgements}
I appreciate discussions with A. Geim, K. Novoselov, T. Stauber, E. Prada, P. San-Jos\'e, and I. Zozoulenko.
This work has been financially supported by MICINN of Spain under Grants Nos. MAT07-67845 and CONSOLIDER CSD2007-00010.
\end{acknowledgements}


\begin{thebibliography}{30}
\bibitem{footnote} Although the exact range of validity of Eq. \ref{scaling} 
is difficult to ascertain, one possibility is to            
compute the dimensionless parameter $K_0$ (see, e.g., Ref. \onlinecite{lewenkopf:081410}), 
which encondes the strength of the disorder for long range models. We have obtained values 
in the range 1-40 which places us, for strong disorder, well into the diffusive regime for scales as short as $L=4$nm
and, for weak disorder, in the quasi-ballistic regime  (see Ref. \onlinecite{bardarson:106801}).
\end{thebibliography}

\end{document}